\documentclass[aps,showpacs,prd,twocolumn]{revtex4}
\usepackage{graphicx}
\usepackage{amssymb}
\usepackage{amsmath}

\newcommand{\be}{\begin{equation}}
\newcommand{\ee}{\end{equation}}
\newcommand{\ben}{\begin{eqnarray}}
\newcommand{\een}{\end{eqnarray}}
\newcommand{\bes}{\begin{subequations}}
\newcommand{\ees}{\end{subequations}}

\begin{document}

\title{BPS Solutions to a Generalized Maxwell-Higgs Model}
\author{D. Bazeia$^{1,2}$, E. da Hora$^{1,3}$, C. dos Santos$^{4}$ and R.
Menezes$^{2,5}$.}
\affiliation{$^{1}${Departamento de F\'{\i}sica, Universidade Federal da Para\'{\i}ba,
58051-970, Jo\~{a}o Pessoa, Para\'{\i}ba, Brazil.}\\
$^{2}${Departamento de F\'{\i}sica, Universidade Federal da Campina Grande,
58109-970, Campina Grande, Para\'{\i}ba, Brazil.}\\
$^{3}$Department of Mathematical Sciences, Durham University, DH1 3LE,
Durham, County Durham, United Kingdom.\\
$^{4}${Centro de F\'{\i}sica e Departamento de F\'{\i}sica e Astronomia,
Faculdade de Ci\^{e}ncias da Universidade do Porto, 4169-007, Porto,
Portugal.}\\
$^{5}${Departamento de Ci\^{e}ncias Exatas, Universidade Federal da Para%
\'{\i}ba, 58297-000, Rio Tinto, Para\'{\i}ba, Brazil.}}

\begin{abstract}
We look for topological BPS solutions of an Abelian-Maxwell-Higgs theory
endowed by non-standard kinetic terms to both gauge and scalar fields. Here,
the non-usual dynamics are controlled by two positive functions, $G\left(
\left\vert \phi \right\vert \right) $ and $w\left( \left\vert \phi
\right\vert \right) $, which are related to the self-dual scalar potential $%
V\left( \left\vert \phi \right\vert \right) $ of the model by a fundamental
constraint. The numerical results we found present interesting new features,
and contribute to the development of the recent issue concerning the study
of generalized models and their applications.
\end{abstract}

\maketitle


\section{Introduction}

In the context of classical field theories, topological structures, such as
kinks \cite{n0}, vortices \cite{n1} and magnetic monopoles \cite{n3}, are
described as static solutions to some nonlinear models. In particular, such
models must allow for the spontaneous symmetry breaking mechanism, since
topological solutions are formed during a symmetry breaking phase
transition. In this sense, topologically non-trivial configurations are of
great interesting to physics \cite{n5}, mainly concerning the cosmological
consequences they may engender, since such configurations can appear in a
rather natural way, during phase transitions in the early universe.

In particular, vortices are described as rotationally symmetric solutions of
a planar Abelian-Maxwell-Higgs model endowed by a fourth-order scalar
potential for the matter self-interaction, which also introduces symmetry
breaking and nonlinearity. In the usual case, the Maxwell term controls the
dynamics of the gauge field, and the covariant derivative squared term
controls the dynamics of the scalar field. In such context, vortices are
finite-energy solutions of a set of two coupled first-order differential
equations, named Bogomol'nyi-Prasad-Sommerfield (BPS) equations \cite{Bogo}.
In this case, the BPS vortices are the minimal-energy solutions of the
model, and they have interesting applications, mainly concerning the
superconductivity phenomena \cite{n1} and the superfluid He$^{4}$.

During the last years, beyond the standard configurations previously cited,
modified ones, also named \textit{topological k-solutions}, have been
intensively studied. These solutions arise in a special class of theoretical
field models, generically named \textit{k-theories}, which are endowed by
non-usual kinetic terms. As expected, such terms change the dynamics of the
overall model under investigation. Moreover, it is important to reinforce
that the idea of a non-standard dynamics arises in a rather natural way, in
the context of string theories.

In general, k-theories have been used as effective models mainly in the
cosmological scenario. Here, many authors have used the so-called k-essence
models \cite{n7h} to investigate the present accelerated inflationary phase
of the universe \cite{n8}. Also, such models can be used to study strong
gravitational waves \cite{n13}, dark matter \cite{n11}, tachyon matter \cite%
{n10} and others \cite{n14}. There are also important motivations concerning
the strong interaction physics; see, for instance, Ref.~\cite{n7a}.

In this context, the overall conclusion is that the existence of topological
structures is quite sensible to the use of non-standard kinetic terms. Here,
the more interesting issue is that such structures can exist even as
solutions of some theories which do not allow for the spontaneous symmetry
breaking mechanism \cite{n14a}. On the other hand, a rather natural way to
study such topological solutions is comparing them with their canonical
counterparts. In this sense, some of us have already investigated
topological solutions in the context of k-field models endowed by
spontaneous symmetry breaking potentials for the scalar-matter
self-interaction, and interesting results can be found, for instance, in Ref.~\cite{n14b}.
Other interesting results concerning k-field models and their
classical solutions can be found, for instance, in Ref.~\cite{n14c}.

Another interesting issue concerning topological k-solutions is that they
can be either much larger or much smaller than their usual counterparts. In
this sense, k-bosons can mediate either large-range or small-range
interactions. Also, important physical quantities, such as energy density
and electric and magnetic fields, can exhibit proeminent variations on their
tipical profiles, including on their maximum values; for a detailed
treatment of some of such features, see Ref.~\cite{n15}.

In the present context, some k-theories can also support topological
solutions with a finite wavelength. These solutions, generically named 
\textit{compactons} \cite{n16}, are quite different from the standard
topological structures, which interact even if separated by an infinite
displacement, since they have an infinite wavelength: two adjacent
compactons will interact only if they come into close contact, due to their
already explained finite wavelength. In this sense, compactons are most
appropriated to describe particle-like configurations than the usual
non-compactified topological structures.

So, in this work we present new results concerning the topological solutions
of an Abelian-Maxwell-Higgs model endowed by non-usual kinetic terms to both
gauge and scalar fields. Here, the non-standard dynamics are introduced by
two positive functions, $G\left( \left\vert \phi \right\vert \right) $ and $%
w\left( \left\vert \phi \right\vert \right) $, which couple with the Maxwell
term and with the covariant derivative squared term, respectively. The
Euler-Lagrange equations of motion of such theory are hard to solve, then we
focus our attention only on the finite-energy solutions of the BPS equations
of the model. These equations can be obtained by minimizing the energy
functional of the model, which can be done via an important constraint
between $G\left( \left\vert \phi \right\vert \right) $, $w\left( \left\vert
\phi \right\vert \right) $ and the scalar potential $V\left( \left\vert \phi
\right\vert \right) $ for the matter self-interaction; see eqs.\eqref{ed1}
and \eqref{ir} below. In this case, rotationally symmetric BPS vortices are
described as the minimal-energy solutions of the cited model, and they
engender interesting features, as explained below.

This paper is outlined as follows: in the next Sec.\ref{general}, we
introduce the model and develop the theoretical framework which allows us to
get to its Bogomol'nyi-Prasad-Sommerfield equations. In Sec.\ref{numerical},
we prove the consistence of the theoretical framework previously developed
by using it to investigate the existence of new BPS\ states. Here, we note
that such states are constrained to the choices made for $G\left( \left\vert
\phi \right\vert \right) $ and $w\left( \left\vert \phi \right\vert \right) $%
: for any acceptable pair of such functions, there is a corresponding BPS\
configuration. Also in Sec.\ref{numerical}, we show how to use the
theoretical framework presented in this work to recover the recover the BPS\
results concerning the standard Maxwell-Higgs model. In Section \ref%
{numerical2}, we perform the numerical analysis concerning the new BPS\
states previously presented, we depict the corresponding minimal-energy
modified solutions and comment on their main features. Finally, in Sec.\ref%
{end}, we present our conclusions and perspectives.

From now on, we use standard conventions, including natural units system,
and a plus-minus signature for the planar Minkowski metric: diag$\left( \eta
^{\mu \nu }\right) =\left( +--\right) $.


\section{The model}

\label{general}

In this section, we introduce the model. It is described by the $\left(
2+1\right) $-dimensional Lagrange density%
\begin{equation}
\mathcal{L}_{G}=-\frac{1}{4}G\left( \left\vert \phi \right\vert \right)
F_{\mu \nu }F^{\mu \nu }+w\left( \left\vert \phi \right\vert \right)
\left\vert D_{\mu }\phi \right\vert ^{2}-V\left( \left\vert \phi \right\vert
\right) \text{ .}  \label{l1}
\end{equation}%
Here, $F_{\mu \nu }=\partial _{\mu }A_{\nu }-\partial _{\nu }A_{\mu }$ is
the usual Faraday field strength tensor, $D_{\mu }\phi =\partial _{\mu }\phi
+ieA_{\mu }\phi $ is the covariant derivative and $V\left( \left\vert \phi
\right\vert \right) $ is the spontaneous symmetry breaking potential. Also, $%
G\left( \left\vert \phi \right\vert \right) $ is the "dieletric function",
and $w\left( \left\vert \phi \right\vert \right) \left\vert D_{\mu }\phi
\right\vert ^{2}$ stands for the non-standard kinetic matter term. Both $%
G\left( \left\vert \phi \right\vert \right) $ and $w\left( \left\vert \phi
\right\vert \right) $ are dimensionless functions to be given below, as
functions of the amplitude of the scalar field.

In the case of vortices, it is convenient to deal with dimensionless
variables. So, for simplicity, we introduce the mass scale $M$, and use it
to implement the scale transformations: $x^{\mu }\rightarrow x^{\mu }/M$, $%
\phi \rightarrow M^{1/2}\phi $, $A^{\mu }\rightarrow M^{1/2}A^{\mu }$, $%
e\rightarrow M^{1/2}e$ and $\upsilon \rightarrow M^{1/2}\upsilon $, where $%
\upsilon $ stands for the symmetry breaking parameter of the model. In this
case, we get $\mathcal{L}_{G}\rightarrow M^{3}\mathcal{L}$, with $\mathcal{L}
$ being the dimensionless Lagrange density to be used from now on, which has
the same functional form of $\mathcal{L}_{G}$. Moreover, we omit the
coupling constant related to the scalar matter self-interaction. Also, we
take $e$ and $\upsilon $ as real and positive parameters.

To search for vortex solutions, the canonical procedure is to deal with the
Euler-Lagrange equations of the model. In the present case, these equations
are%
\begin{equation}
G\partial _{\mu }F^{\mu \lambda }+F^{\mu \lambda }\partial _{\mu
}G=J^{\lambda }\text{ ,}  \label{mele1}
\end{equation}%
and%
\begin{eqnarray}
&&\left. w\partial _{\mu }\partial ^{\mu }\left\vert \phi \right\vert
+\partial _{\mu }\left\vert \phi \right\vert \partial ^{\mu }w-e^{2}A_{\mu
}A^{\mu }\left\vert \phi \right\vert w\right.  \notag \\
&&\left. =\frac{1}{2}\left\vert D_{\mu }\phi \right\vert ^{2}\frac{dw}{%
d\left\vert \phi \right\vert }-\frac{F^{2}}{8}\frac{dG}{d\left\vert \phi
\right\vert }-\frac{1}{2}\frac{dV}{d\left\vert \phi \right\vert }\text{ .}%
\right.  \label{mele12}
\end{eqnarray}%
Here, we take $F^{2}=F_{\mu \nu }F^{\mu \nu }$ and%
\begin{equation}
J^{\mu }=-2e^{2}w\left\vert \phi \right\vert ^{2}A^{\mu }
\end{equation}%
as the modified 4-current vector.

The Gauss law for time-independent fields can be written as%
\begin{equation}
G\partial _{k}\partial ^{k}A^{0}+\partial _{k}A^{0}\partial
^{k}G=-2e^{2}w\left\vert \phi \right\vert ^{2}A^{0}\text{ ,}  \label{ngl}
\end{equation}%
where $\mathbf{E}=-\overrightarrow{\nabla }A^{0}$. We note that the 
Eq.~\eqref{ngl} is trivially verified by $A^{0}=0$. So, we fix this gauge and
use it from now on.

We look for vortex solutions of the form%
\begin{equation}
\phi \left( r,\theta \right) =\upsilon g\left( r\right) e^{in\theta }\text{ ,%
}  \label{va1}
\end{equation}%
\begin{equation}
\mathbf{A}=-\frac{\widehat{\theta }}{er}\left( a\left( r\right) -n\right) 
\text{ .}  \label{va2}
\end{equation}%
Here, $r$ and $\theta $ are polar coordinates, and $n=\pm 1,\pm 2,\pm 3,...$
is the winding number (vorticity) of the solution. In terms of \eqref{va1}
and \eqref{va2}, the Euler-Lagrange equations \eqref{mele1} and %
\eqref{mele12} can be rewritten as%
\begin{equation}
G\frac{d^{2}a}{dr^{2}}+\left( \frac{dG}{dr}-\frac{G}{r}\right) \frac{da}{dr}%
=2e^{2}\upsilon ^{2}g^{2}aw\text{ ,}  \label{el1}
\end{equation}%
\begin{eqnarray}
&&\left. w\left( \frac{d^{2}g}{dr^{2}}+\frac{1}{r}\frac{dg}{dr}-\frac{a^{2}g%
}{r^{2}}\right) -\frac{1}{4\upsilon ^{2}}\left( \frac{1}{er}\frac{da}{dr}%
\right) ^{2}\frac{dG}{dg}\right.  \notag \\
&&\left. \text{ \ \ }=\frac{1}{2\upsilon ^{2}}\frac{dV}{dg}-\frac{1}{2}%
\left( \left( \frac{dg}{dr}\right) ^{2}-\frac{g^{2}a^{2}}{r^{2}}\right) 
\frac{dw}{dg}\text{ .}\right.  \label{el2}
\end{eqnarray}%
Equations \eqref{el1} and \eqref{el2} are the Euler-Lagrange equations of
motion to the profile functions $a\left( r\right) $ and $g\left( r\right) $,
respectively.

To solve \eqref{el1} and \eqref{el2}, we need to specify the model. In
general, we can do it by choosing non-trivial forms to $G\left( g\right) $
and $w\left( g\right) $. In this case, we have to keep in mind that both
these functions must be positive, in order to avoid problems with the energy
of the model; see the expression for the energy density \eqref{ed1} below.
Also, we need to choose a Higgs potential $V\left( g\right) $ which allows
for the spontaneous symmetry breaking mechanism.

Before that, we note that the limit $w\left( g\right) \rightarrow 1$ leads
us back to the model studied in \cite{db}, which is supported by
applications concerning the interaction between quarks and gluons \cite{db2}%
. In this case, the limit $G\left( g\right) \rightarrow 1$ leads us back to
the usual Maxwell-Higgs theory. In this context, if we choose the symmetry
breaking Higgs potential%
\begin{equation}
V_{s}\left( g\right) =\frac{e^{2}\upsilon ^{4}}{2}\left( 1-g^{2}\right) ^{2}%
\text{ ,}  \label{up}
\end{equation}%
the Euler-Lagrange equations \eqref{el1} and \eqref{el2} can then be
rewritten as%
\begin{equation}
\frac{d^{2}a}{dr^{2}}-\frac{1}{r}\frac{da}{dr}=2e^{2}\upsilon ^{2}g^{2}a%
\text{ ,}  \label{el3}
\end{equation}%
\begin{equation}
\frac{d^{2}g}{dr^{2}}+\frac{1}{r}\frac{dg}{dr}-\frac{a^{2}g}{r^{2}}%
=e^{2}\upsilon ^{2}g\left( g^{2}-1\right) \text{ .}  \label{el4}
\end{equation}%
According to our conventions, eqs.\eqref{el3} and \eqref{el4} are completely
solvable by the first order differential equations%
\begin{equation}
\frac{dg}{dr}=\pm \frac{ga}{r}\text{ ,}  \label{bps1}
\end{equation}%
\begin{equation}
\frac{1}{r}\frac{da}{dr}=\pm e^{2}\upsilon ^{2}\left( g^{2}-1\right) \text{ .%
}  \label{bps2}
\end{equation}

The solutions of eqs.\eqref{bps1} and \eqref{bps2} are the
Bogomol'nyi-Prasad-Sommerfield (BPS) states of the standard Maxwell-Higgs
model \eqref{up}. These solutions are the well-known
Abrikosov-Nielsen-Olesen (ANO) ones \cite{n1}, which solve the equations of
motion \eqref{el3} and \eqref{el4} by minimazing the energy of the resulting
vortex configurations.

In general, for non-trivial choices to $G\left( g\right) $ and $w\left(
g\right) $, the equations of motion \eqref{el1} and \eqref{el2} will be much
more sofisticated than the eqs.\eqref{el3} and \eqref{el4}. So, in order to
get an useful insight about the non-trivial case, we consider the expression
for the energy density of the modified vortex solutions:%
\begin{equation}
\varepsilon =\frac{G}{2e^{2}}\left( \frac{1}{r}\frac{da}{dr}\right)
^{2}+\upsilon ^{2}w\left( \left( \frac{dg}{dr}\right) ^{2}+\frac{g^{2}a^{2}}{%
r^{2}}\right) +V\text{ .}  \label{ed1}
\end{equation}%
From Eq.~\eqref{ed1}, we note that the presence of non-trivial $G\left(
g\right) $ and $w\left( g\right) $ makes it hard to obtain the BPS states of
the modified model. Even in this case, the existence of such states is still
possible, and it is closely related to an important constraint between $%
G\left( g\right) $, $w\left( g\right) $ and $V\left( g\right) $:%
\begin{equation}
\frac{d}{dg}\sqrt{GV}=\sqrt{2}e\upsilon ^{2}wg\text{ .}  \label{ir}
\end{equation}%
As we clarify below, for any positive choices to $G\left( g\right) $ and $%
w\left( g\right) $, there is a corresponding symmetry breaking Higgs
potential $V\left( g\right) $ as a solution of \eqref{ir}. In this context,
generalized first order equations can be found by minimizing the energy
functional \eqref{ed1}.

To search for BPS\ vortex solutions in the modified model, we need to know
how the functions $g\left( r\right) $ and $a\left( r\right) $ behave, near
the origin and asymptotically. Near the origin, these functions must avoid
singular fields. So, they have to behave according to%
\begin{equation}
g\left( r\rightarrow 0\right) \rightarrow 0\text{ \ \ and \ \ }a\left(
r\rightarrow 0\right) \rightarrow n=1\text{ ,}  \label{bc1}
\end{equation}%
where we have fixed $n=1$, for simplicity. Also, the symmetry breaking
vortex configurations must have a finite total energy. So, as a condition to
make the energy finite, the energy density \eqref{ed1} must vanish. Then,
asymptotically, $g\left( r\right) $ and $a\left( r\right) $ have to obey%
\begin{equation}
g\left( r\rightarrow \infty \right) \rightarrow 1\text{ \ \ and \ \ }a\left(
r\rightarrow \infty \right) \rightarrow 0\text{ .}  \label{bc2}
\end{equation}

In the next Sec. III, we use the theoretical framework developed in this
Section to investigate the existence of BPS states in the generalized model.
Also, we present and comment the resulting numerical solutions. Finally, we
show how to map the standard Maxwell-Higgs and Chern-Simons-Higgs first
order equations; even in these cases, Eq.~\eqref{ir} leads us to a physically
different rotationally symmetric solutions.


\section{New BPS states}

\label{numerical}

We now pay due attention to the modified BPS states themselves. Here, we
choose positive functional forms to $G\left( g\right) $ and $w\left(
g\right) $. Then, we solve the constraint \eqref{ir} to get to the
consistent Higgs potential $V\left( g\right) $, which must allow for the
spontaneous symmetry breaking mechanism. A posteriori, we use these
conventions to obtain the first order differential equations of the modified
model, by minimizing its energy functional \eqref{ed1}. Then, in the next
Sec.IV, we numerically solve these equations, according to the boundary
conditions \eqref{bc1} and \eqref{bc2}. Finally, in Sec.V, we comment the
main features of the resulting solutions.

The model to be studied here has two unusual functions, $G\left( g\right) $
and $w\left( g\right) $, which we have to fix to determine how they affect
the vortex solutions. The function $G\left( g\right) $ stands for a
"dieletric function", and it controls the non-standard kinetic term to the
gauge field. On the other hand, the function $w\left( g\right) $ controls
the non-usual kinetic scalar matter term. Both these functions are
dimensionless, and are functions of the amplitude of the scalar field. Also,
they must be positive, in order to avoid problems with the energy of the
model; see Eq.~\eqref{ed1}.

A rather natural way to investigate the consistence of the theoretical
framework developed in the previous Sec. II is to study the BPS states of
the standard Maxwell-Higgs model, which is defined by $G\left( g\right)
=w\left( g\right) =1$. In this case, Eq.~\eqref{ir} leads us to the potential %
\eqref{up}. Then, the energy density \eqref{ed1} can be rewritten as%
\begin{eqnarray}
\varepsilon &=&\frac{1}{2}\left( \frac{1}{er}\frac{da}{dr}\mp e\upsilon
^{2}\left( g^{2}-1\right) \right) ^{2}+\upsilon ^{2}\left( \frac{dg}{dr}\mp 
\frac{ga}{r}\right) ^{2}  \notag \\
&&\mp \frac{\upsilon ^{2}}{r}\frac{da}{dr}\pm \frac{\upsilon ^{2}}{r}\frac{d%
}{dr}\left( g^{2}a\right) \text{ .}  \label{se}
\end{eqnarray}%
The resulting total energy $E$ is minimized by the first order equations %
\eqref{bps1} and \eqref{bps2}. In this context, the energy of the BPS states
is%
\begin{equation}
E_{s}=\int \varepsilon \left( r\right) d^{2}r=2\pi \upsilon ^{2}\left\vert
n\right\vert \text{ .}  \label{se1}
\end{equation}%
As usual, this is the lower bound of the energy functional, i.e., the
Bogomol'nyi bound. Also, it is quantized according to the winding number $n$.

Now, we introduce an interesting new model, which is defined by%
\begin{equation}
G\left( g\right) =\frac{\left( g^{2}+3\right) ^{2}}{g^{2}}\text{ \ \ and \ \ 
}w\left( g\right) =2\left( g^{2}+1\right) \text{ .}  \label{nm}
\end{equation}%
According to these choices, we solve \eqref{ir} to get to the potential%
\begin{equation}
V\left( g\right) =g^{2}V_{s}\left( g\right) \text{ ,}  \label{nmp}
\end{equation}%
which allows for the spontaneous symmetry breaking mechanism, as desired. We
point out that this new model is not a parametrization of the standard
Maxwell-Higgs one, since the vacuum manifold of the two models are quite
different: it is a circle for the usual model \eqref{up}, while it is a dot
surrounded by a circle for the new model \eqref{nmp}.

Using eqs.\eqref{nm} and \eqref{nmp}, the energy density \eqref{ed1} can
then be rewritten as%
\begin{eqnarray}
\varepsilon &=&\frac{\left( g^{2}+3\right) ^{2}}{2g^{2}}\left( \frac{1}{er}%
\frac{da}{dr}\mp \frac{e\upsilon ^{2}g^{2}\left( g^{2}-1\right) }{\left(
g^{2}+3\right) }\right) ^{2}  \notag \\
&&+2\upsilon ^{2}\left( g^{2}+1\right) \left( \frac{dg}{dr}\mp \frac{ga}{r}%
\right) ^{2}  \notag \\
&&\pm \frac{\upsilon ^{2}}{r}\frac{d}{dr}\left( a\left( g^{2}-1\right)
\left( g^{2}+3\right) \right) \text{ .}  \label{bpse1}
\end{eqnarray}%
Then, the corresponding energy functional is minimized by the equations%
\begin{equation}
\frac{dg}{dr}=\pm \frac{ga}{r}\text{ ,}  \label{bpsg1}
\end{equation}%
\begin{equation}
\frac{1}{r}\frac{da}{dr}=\pm \frac{e^{2}\upsilon ^{2}g^{2}\left(
g^{2}-1\right) }{\left( g^{2}+3\right) }\text{ .}  \label{bpsg2}
\end{equation}%
We note that the modified equations \eqref{bpsg1} and \eqref{bpsg2} can not
be parametrized into the standard ones, i.e., eqs.\eqref{bps1} and %
\eqref{bps2}. Even in this case, the energy of the modified field solutions
is bounded from below, and the Bogomol'nyi bound is%
\begin{equation}
E=3E_{s}\text{ .}
\end{equation}%
Here, as in the standard case, the total energy of the BPS states is
quantized; see Eq.~\eqref{se1}.

In order to reinfoce the consistence of the theoretical framework presented
in this work, we introduce another modified model. It is defined by%
\begin{equation}
G\left( g\right) =\left( g^{2}+1\right) ^{2}\text{ \ \ and \ \ }w\left(
g\right) =2g^{2}\text{ .}  \label{nm2}
\end{equation}%
According to these conventions, Eq.~\eqref{ir} leads us to the standard Higgs
potential \eqref{up}. So, the modified model \eqref{nm2} has the same vacuum
structure as the usual model. Even in this context, the new model is not a
parametrization of the standard one; see the BPS\ eqs.\eqref{bpsg3} and %
\eqref{bpsg4} below. The modified energy density can be written in the form%
\begin{eqnarray}
\varepsilon &=&\frac{\left( g^{2}+1\right) ^{2}}{2}\left( \frac{1}{er}\frac{%
da}{dr}\mp \frac{e\upsilon ^{2}\left( g^{2}-1\right) }{\left( g^{2}+1\right) 
}\right) ^{2}  \notag \\
&&+2\upsilon ^{2}g^{2}\left( \frac{dg}{dr}\mp \frac{ga}{r}\right) ^{2} 
\notag \\
&&\pm \frac{\upsilon ^{2}}{r}\frac{d}{dr}\left( a\left( g^{2}-1\right)
\left( g^{2}+1\right) \right) \text{ ,}  \label{bpse2}
\end{eqnarray}%
and the resulting energy functional is minimized by the first order equations%
\begin{equation}
\frac{dg}{dr}=\pm \frac{ga}{r}\text{ ,}  \label{bpsg3}
\end{equation}%
\begin{equation}
\frac{1}{r}\frac{da}{dr}=\pm \frac{e^{2}\upsilon ^{2}\left( g^{2}-1\right) }{%
\left( g^{2}+1\right) }\text{ .}  \label{bpsg4}
\end{equation}%
Then, the Bogolmol'nyi bound is also given by \eqref{se1}, and we conclude
that the modified solutions \eqref{bpsg3} and \eqref{bpsg4} have the same
energy of the standard \eqref{bps1} and \eqref{bps2} ones.

To end this Section, we show how to use the theoretical framework developed
in this work to map the standard Maxwell- and Chern-Simons-Higgs first order
differential equations. We do it by reviewing the model studied in \cite{db}%
, which is defined by $w\left( g\right) =1$. In this case, as presented in
that work, an interesting choice to the dieletric function is%
\begin{equation}
G\left( g\right) =\left( 1+\lambda \right) \left( 1+2\lambda \frac{%
e^{2}\upsilon ^{2}}{k^{2}}g^{2}\right) ^{-1}\text{ .}  \label{um}
\end{equation}%
Here, $\lambda $ is a real auxiliary parameter which controls the model, and 
$k$ stands for the coupling constant corresponding to the Chern-Simons term.
In the limit $\lambda \rightarrow 0$, Eq.~\eqref{um} gives $G\left( g\right)
\rightarrow 1$, and \eqref{ir} reproduces the standard Maxwell-Higgs system %
\eqref{up}. On the other hand, the limit $\lambda \rightarrow \infty $ gives%
\begin{equation}
G\left( g\right) =\frac{k^{2}}{2e^{2}\upsilon ^{2}g^{2}}\text{ .}
\label{csg}
\end{equation}%
We then solve Eq.~\eqref{ir} to get%
\begin{equation}
V\left( g\right) =\frac{e^{4}\upsilon ^{6}}{k^{2}}g^{2}\left( g^{2}-1\right)
^{2}\text{ ,}  \label{csp}
\end{equation}%
which is exactly the sixth-order symmetry breaking Higgs potential one finds
in the usual self-dual Chern-Simons-Higgs model. Now, from eqs.\eqref{csg}
and \eqref{csp}, the energy density \eqref{ed1} can be written in the form%
\begin{eqnarray}
\varepsilon &=&\frac{k^{2}}{4e^{2}\upsilon ^{2}g^{2}}\left( \frac{1}{er}%
\frac{da}{dr}\mp \frac{2e^{3}\upsilon ^{4}}{k^{2}}g^{2}\left( g^{2}-1\right)
\right) ^{2}  \notag \\
&&+\upsilon ^{2}\left( \frac{dg}{dr}\mp \frac{ga}{r}\right) ^{2}\pm \upsilon
^{2}\frac{1}{r}\frac{d}{dr}\left( a\left( g^{2}-1\right) \right) \text{ .}
\end{eqnarray}%
The resulting total energy is minimized by the differential equations%
\begin{equation}
\frac{dg}{dr}=\pm \frac{ga}{r}\text{ ,}  \label{cs1}
\end{equation}%
\begin{equation}
\frac{1}{r}\frac{da}{dr}=\pm \frac{2e^{4}\upsilon ^{4}}{k^{2}}g^{2}\left(
g^{2}-1\right) \text{ ,}  \label{cs2}
\end{equation}%
and the Bogomol'nyi bound is $E=E_{s}$. The equations \eqref{cs1} and %
\eqref{cs2} are just the BPS\ ones for the standard Chern-Simons-Higgs
system \eqref{csp}. In this sense, Eq.~\eqref{um} works as a unified way to
map both the Maxwell- and the Chern-Simons-Higgs first order differential
equations. Finally, we point out that, despite the possibility of
reproducing such Chern-Simons equations, the unified picture \eqref{um}
leads us to a physically different situation, since the modified model we
consider here only supports non-charged field solutions; see the Gauss law %
\eqref{ngl} above. In this case, the modified nontopological field solutions
are unstable, and they can decay into the elementary mesons of the model.
This fact makes an important difference, since the nontopological
configurations presented by the self-dual Chern-Simons-Higgs theory are
stable, and can not decay.

In the next Section, we present the modified first order\ numerical
solutions for the profile funtions $g\left( r\right) $ and $a\left( r\right) 
$. Also, we comment on the main features of such solutions.


\section{Numerical solutions}

\label{numerical2}

Let us focus attention on the modified numerical solutions themselves. The
equations to be solved are the first order ones \eqref{bpsg1} and %
\eqref{bpsg2}, and \eqref{bpsg3} and \eqref{bpsg4}. Also, we solve the
standard equations \eqref{bps1} and \eqref{bps2}, for comparison. In all
cases, the functions $g\left( r\right) $ and $a\left( r\right) $ must obey
the boundary conditions \eqref{bc1} and \eqref{bc2}.

In the present work, the numerical strategy to be employed is the relaxation
one. In this case, it is necessary to input an approximated field solution.
Then, our algorithm will "relax" it into the correct one. To start the
numerical analysis, we consider a variation of the standard Maxwell-Higgs
model, in which the potential for the self-interaction for the scalar field
is not present. Then, we use its solutions to solve the self-dual
Maxwell-Higgs case, i.e., the equations \eqref{bps1} and \eqref{bps2}, from
which we get the well-understood Abrikosov-Nielsen-Olesen (ANO)
configurations. We then use these solutions to initialize the numerical
study of the modified theory.

Starting from such ANO configurations, we numerically solve the modified BPS
equations presented in the previous Section, i.e., eqs.\eqref{bpsg1} and %
\eqref{bpsg2}, and also eqs.\eqref{bpsg3} and \eqref{bpsg4}, for $e=\upsilon
=n=1$. The solutions for the profile functions $g\left( r\right) $ and $%
a\left( r\right) $ are plotted in Figs. 1 and 2, respectively. Also, we
depict the solutions for the corresponding energy densities; see Figure 3.

\begin{figure}[tbph]
\centering
\includegraphics[width=8.5cm]{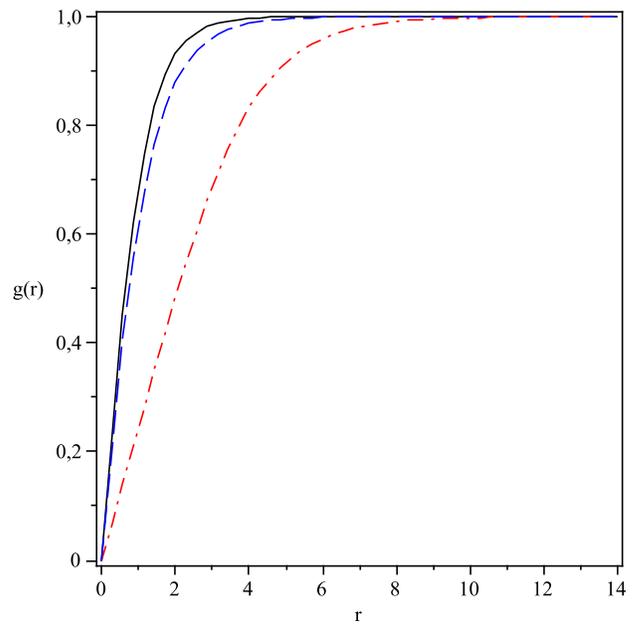}
\par
\vspace{-0.3cm}
\caption{Numerical BPS solutions to $g\left( r\right) $ for the models 
\eqref{nm} (dash-dotted red line) and \eqref{nm2} (dashed blue line). The
standard solution is also depicted (solid black line), for comparison.}
\end{figure}


In Figure 1, we present the solutions for the profile function $g\left(
r\right) $, and we see that both the modified profiles reach their vacuum
values more slowly than their canonical counterpart. In this sense, such
solutions have a core which is greater than the usual Maxwell-Higgs one.
Here, the conclusion is that, in general, the introduction of a
non-canonical dynamics allows for the existence of self-dual field solutions 
$g\left( r\right) $ with a increased characteristic length. Also, we note
that the solution for \eqref{bpsg1} and \eqref{bpsg2} goes to its vacuum
configuration more slowly than that for \eqref{bpsg3} and \eqref{bpsg4}. So,
beyond the fact of increasing the core of the solutions, we note that %
\eqref{nm} increases it more than the \eqref{nm2}. We believe that this fact
is related with the corresponding self-dual Higgs potential, which is of
sixth-order for the model \eqref{nm}, and of fourth-order for the \eqref{nm2}
one; see equations \eqref{nmp} and \eqref{up} above.

\begin{figure}[tbph]
\centering\includegraphics[width=8.5cm]{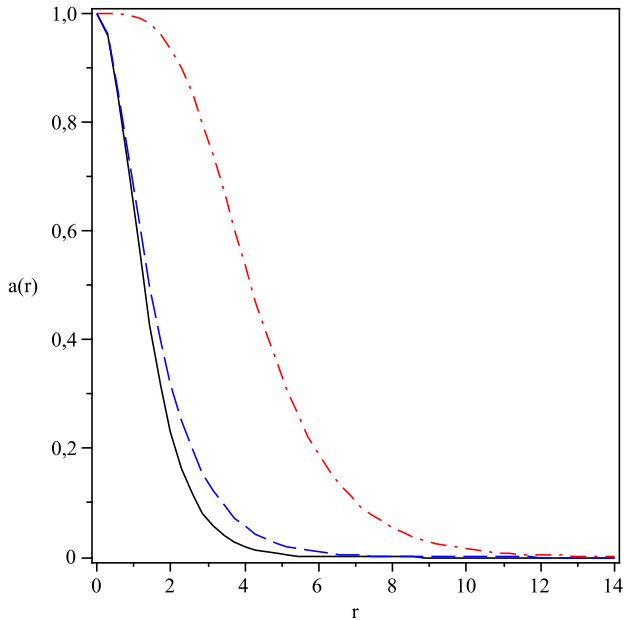}
\par
\vspace{-0.3cm}
\caption{Numerical BPS solutions to $a\left( r\right) $. Conventions as in
FIG. 1.}
\end{figure}


In Fig. 2, we depict the numerical solutions for the function $a\left(
r\right) $. Here, as in the Fig. 1, we note that the modified solutions go
to their vacuum states more slowly than the standard Maxwell-Higgs profile,
and so such solutions have a increased characteristic length. This behaviour
reinforces the previous conclusion, which states that a non-standard
dynamics leads to a self-dual field solutions with a increased core. We also
see that, beyond the fact of increasing the core of $a\left( r\right) $, the
model \eqref{nm} enlarges it more than the \eqref{nm2}. Finally, we point
out the existence of a proeminent plateau in the profile related to %
\eqref{nm}, near the origin: such structure is also present in the self-dual
Chern-Simons-Higgs case, which is governed by the potential \eqref{csp}. In
this context, it is interesting to note that the existence of such
proeminent plateau seems to be closely related to the vacuum manifold of a
sixth-order symmetry breaking potential, since it is also present in the
modified model \eqref{nm}, which is governed by \eqref{nmp}.

The Fig. 3 encloses the solutions for the energy densities of the new BPS
states; see equations \eqref{bpse1} and \eqref{bpse2}. We see that the
profile for the energy density corresponding to the modified model \eqref{nm}
is quite different from that of the standard Maxwell-Higgs theory: in the
canonical case, the energy density reaches its maximum value in the limit $%
r\rightarrow 0$, and it is monotonically decrescent for all $r$. On the
other hand, in the modified case \eqref{nm}, the energy density reaches its
maximum value at some finite distance $R$ from the origin, and it is not
monotonically decrescent for all values of the independent variable. In this
context, this difference reinforces our previous conclusion, since such
modified behaviour mimics the one of the energy density of the usual
self-dual Chern-Simons theory.

\begin{figure}[tbph]
\centering\includegraphics[width=8.5cm]{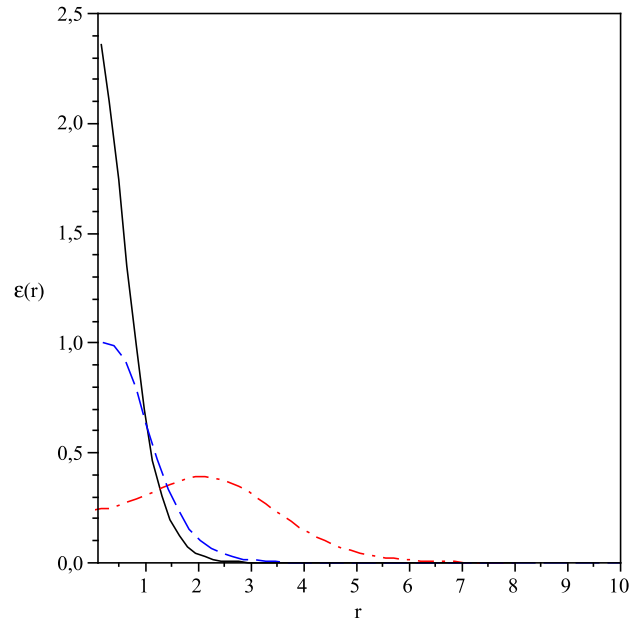}
\par
\vspace{-0.3cm}
\caption{Plots of the energy density $\protect\varepsilon \left( r\right) $.
Conventions as in FIG. 1.}
\end{figure}


The solution for the BPS energy density of the modified model \eqref{nm2} is
also depicted in Figure 3, and we see that its behaviour is qualitatively
similar to that of the Maxwell-Higgs model: it reaches its maximum value as $%
r$ goes to $0$, and it is monotonically decrescent for all $r$. Here,
however, there are two important differences: the first one is on the
maximum value of the modified profile itself, which is smaller than the
usual one, and the second one is on the characteristic length of the
non-standard solution, which is greater than its canonical counterpart.

An important issue concerning the study of topological structures is the
computation of their topological charges, which must be conserved. In the
present case, i.e., for electrically non-charged field solutions of the form %
\eqref{va1} and \eqref{va2}, this charge is related to the magnetic field
generated by such solutions themselves. To investigate this issue, we
introduce the topological current%
\begin{equation}
J^{\mu }=\epsilon ^{\mu \nu \lambda }\partial _{\nu }A_{\lambda }\text{ ,}
\end{equation}%
which is clearly conserved. Here, $\epsilon ^{012}=-1$. The 0$^{\text{th}}$%
-component of such current, that is, the topological charge density, can be
written as%
\begin{equation}
J^{0}=\partial _{x}A_{y}-\partial _{y}A_{x}=B\text{ ,}
\end{equation}%
and we see that it is directly proportional to the magnetic field $B$. In
this case, the topological charge is given by%
\begin{equation}
Q_{T}=\int Bd^{2}r\equiv \Phi _{B}\text{ .}  \label{x1}
\end{equation}%
Here, $\Phi _{B}$\ stands for the flux of the magnetic field.

According \eqref{va2}, \eqref{bc1} and \eqref{bc2}, the magnetic flux $\Phi
_{B}$\ can be rewritten in the form%
\begin{equation}
\Phi _{B}=\frac{2\pi n}{e}\text{ ,}  \label{x2}
\end{equation}%
and we note that the topological charge $Q_{T}$ is conserved, and it is
quantized according to the winding number $n$.

In this sense, to better specify the field configurations studied in this
work, a rather natural way is to consider their corresponding magnetic
fields. In the present context, this field is given by%
\begin{equation}
B=-\frac{1}{r}\frac{da}{dr}\text{ .}  \label{mf}
\end{equation}

The modified numerical profiles for the magnetic field \eqref{mf} are
plotted in Figure 4. Also for such field, the non-usual dynamics introduced
earlier leads to BPS solutions with a increased core, since the modified
profiles have a characteristic length which is greater than the standard
one. In this case, as for the energy densities previously depicted, the
magnetic field associated to \eqref{nm} is quite different from the
canonical self-dual one, which reaches its maximum value for $r\rightarrow 0$%
, and is monotonically decrescent ever: the non-standard solution reaches
its maximum value at some finite distance from the origin, and it is not
monotonically decrescent for all $r$. Such behaviour is similar to that of
the self-dual magnetic field related to the canonical Chern-Simons-Higgs
model.

\begin{figure}[tbph]
\centering\includegraphics[width=8.5cm]{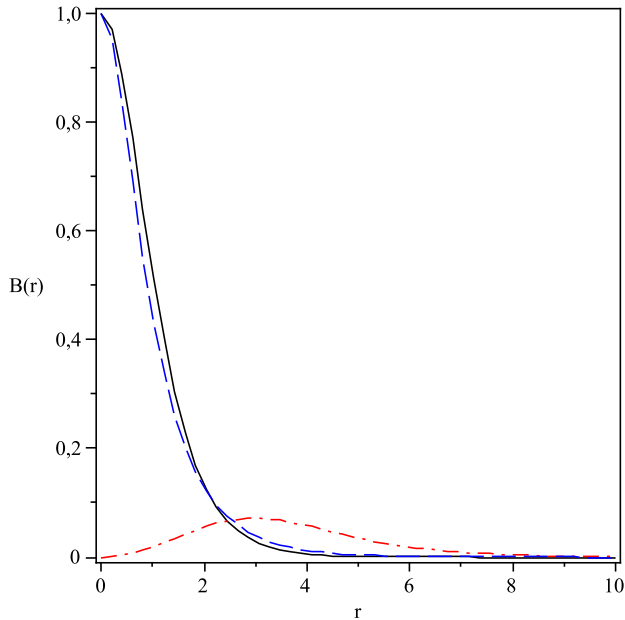}
\par
\vspace{-0.3cm}
\caption{Plots of the magnetic field $B\left( r\right) $. Conventions as in
FIG. 1.}
\end{figure}


To end this Section, we discuss the solution for the magnetic field %
\eqref{mf} related to the model \eqref{nm2}. This solution is also depicted
in Fig. 4, from which we note that the behaviour of such magnetic field is
quite similar to that of the usual Maxwell-Higgs case, since it reaches its
maximum value as $r\rightarrow 0$, and it is monotonically decrescent ever.
Here, we point out the behaviour of the modified solution, which assures the
conservation of the topological charge; see eqs.\eqref{x1} and \eqref{x2}.


\section{Ending comments}

\label{end}

In the present letter, we have considered the existence of rotationally
symmetric BPS solutions in a $\left( 2+1\right) $-dimensional space-time. We
have investigated such solutions in a modified self-dual Maxwell-Higgs model
endowed by a non-standard dynamics. Here, the modification was introduced in
terms of two non-trivial functions, $G\left( \left\vert \phi \right\vert
\right) $ and $w\left( \left\vert \phi \right\vert \right) $, which must be
positive, in order to avoid problems with the energy density \eqref{ed1} of
the model. So, while $G\left( \left\vert \phi \right\vert \right) $ couples
with the Maxwell term and, as a consequence, changes the dynamics of the
gauge field in a non-usual way, $w\left( \left\vert \phi \right\vert \right) 
$ couples with the squared covariant derivative of the non-charged scalar
field, then leading to a non-standard dynamics to such field. In this
context, consistent first order equations were obtained since the
non-trivial functions $G\left( \left\vert \phi \right\vert \right) $ and $%
w\left( \left\vert \phi \right\vert \right) $ are constrained to the
symmetry breaking Higgs potential $V\left( \left\vert \phi \right\vert
\right) $ of the modified model; see Eq.~\eqref{ir}.

We have integrated the BPS equations by means of the relaxation method, and
the numerical results we found are depicted in Figs. 1 and 2, for some
interesting choices to the functions $G\left( \left\vert \phi \right\vert
\right) $ and $w\left( \left\vert \phi \right\vert \right) $; see eqs.%
\eqref{nm} and \eqref{nm2}. According these solutions, we conclude that both
the profile functions $g\left( r\right) $ and $a\left( r\right) $ are quite
sensible to the choices made for $G\left( \left\vert \phi \right\vert
\right) $ and $w\left( \left\vert \phi \right\vert \right) $. In particular,
it is important to reinforce that each of the models \eqref{nm} and %
\eqref{nm2} is related to a very specific symmetry breaking potential $%
V\left( \left\vert \phi \right\vert \right) $ and, as a consequence, such
models present distinct vacuum manifolds; for details, see eqs.\eqref{up}
and \eqref{nmp}.

Also, using the previous results we found for $g\left( r\right) $ and $%
a\left( r\right) $, we have integrated both the energy density $\varepsilon
\left( r\right) $ and the magnetic field $B\left( r\right) $ for the
modified models \eqref{nm} and \eqref{nm2}, and these solutions are depicted
in Figs. 3 and 4, respectively; see eqs.\eqref{bpse1}, \eqref{bpse2} and %
\eqref{mf}. The numerical analysis reveals that the solutions corresponding
the model \eqref{nm} behave as those predicted by the Chern-Simons-Higgs
theory, which reach their maximum values at some finite distance from the
origin, and are not monotonically decrescent for all $r$. On the other hand,
the solutions corresponding the model \eqref{nm2} behave as the
Maxwell-Higgs ones, since they reach their maximum values as $r\rightarrow 0$%
, and are monotonically decrescent for all values of the independent
variable. So, as a consequence, we conclude that also the energy density $%
\varepsilon \left( r\right) $ and the magnetic field $B\left( r\right) $ are
both sensible to the choices made for $G\left( \left\vert \phi \right\vert
\right) $ and $w\left( \left\vert \phi \right\vert \right) $.

We hope that this work may stimulate subsequent analysis in the field,
concerning mainly the features that the modified solutions engender. Also,
we point out that the variation on the characteristic length of the modified
solutions presented here is closely related to the effects of anisotropy,
which is a feature tipically present in the effective field models used to
describe the behaviour engendered by low-energy condensed matter systems. In
this sense, we point out that such effective models are usually based on the
Lorentz-breaking ideia, since it introduces the issue of anisotropy
explicity. In the context of Lorentz-violating models, such effects where
already studied by some of us; see, for instance, Ref.~\cite{n17}.

A very interesting issue concerns the use of modified models to mimic the
very same solutions engendered by the standard Maxwell-Higgs theory. In
particular, such issue is now under consideration, and we hope to report new
results in a near future.

The authors would like to thank CAPES, CNPq (Brazil) and FCT Project
CERN/FP/116358/2010 (Portugal) for partial financial support. Also, we are
grateful to C. Adam, F. Correia and D. Rubiera-Garcia for useful
discussions. E. da Hora thanks the Department of Mathematical Sciences of
Durham University (United Kingdom), for all their hospitality while doing
this work.

\end{document}